\documentclass[a4paper,11pt]{article}

\usepackage{amsmath}
\usepackage{amsfonts}
\usepackage{amssymb}

\begin{document}

\title{On the experimental foundations of the \\
Maxwell equations}

\author{M.\ Haugan$^{1}$ and C.\ L{\"a}mmerzahl$^{2}$ \\
$^{1}$Department of Physics, Purdue University, 1396 Physics
Building, West Lafayette,
  \\ \hspace*{0.5mm} Indiana 47907-1396, USA \\
  $^{2}$Faculty of Physics, University of Konstanz,
  Fach M674, 78457 Konstanz, Germany}

\maketitle

\begin{abstract}
We begin by reviewing the derivation of generalized Maxwell equations
from an operational definition of the electromagnetic field and the
most basic notions of what constitutes a dynamical field theory.
These equations encompass the familiar Maxwell equations as a
special case but, in other cases, can predict birefringence, charge
non-conservation, wave damping and other effects that the familiar
Maxwell equations do not.  It follows that observational constraints
on such effects can restrict the dynamics of the electromagnetic field
to be very like the familiar Maxwellian dynamics, thus, providing
an empirical foundation for the Maxwell equations.  We discuss
some specific observational results that contribute to that foundation.
\end{abstract}

keywords: Maxwell equations, birefringence, charge
conservation

PACS: 03.50 De, 04.80.-y, 41.20.-q, 03.30.+p

\section{Introduction}

According to special relativity and to metric theories of gravity like
general relativity, the dynamics of the electromagnetic field is
intimately connected with the structure of spacetime.  For example,
light cones can be used to identify the causal structure of spacetime
in these theories, and the dynamics of the electromagnetic field
combines with the quantum mechanics of charged particles to determine
the behavior of atomic clocks and standards of length one uses
to map out spacetime geometry.  This connection between
spacetime structure and the dynamics of the electromagnetic field
clearly motivates the sharpest possible experimental tests of the
validity of the familiar Maxwellian dynamics.

Generalized Maxwell equations provide a context in which to design
and interpret such tests.  Section 2 reviews the derivation of these
equations from an operational definition of the electromagnetic field
and the most basic notions of what constitutes a dynamical field
theory.  Since no geometrical spacetime structure like metric or conmnection  
is presumed
by the derivation the generalized Maxwell equations can predict
birefringence, charge non-conservation, wave damping and other
effects not predicted by the familiar Maxwell equations.  The
familiar equations are, however, a special case of the generalized
ones so that experiments which search for effects like those just
mentioned can determine the extent to which that the dynamics of the
electromagnetic field is or is not compatible with spacetime geometry.
Section 3 discusses astronomical observations whose results constrain
electromagnetic field dynamics that predict finite wave propagation
speeds to be very close to the familiar Maxwellian dynamics.  It also
discusses effects which could be used to test the assumption of
finite wave
propagation speed and the assumption that the electrodynamic field
solves a well-posed Cauchy problem.  Section 4 contains a few
closing remarks.

\section{A derivation of the Maxwell equations}

The mathematical structure of the Maxwell equations have been
analyzed in \cite{PuntigamLaemmerzahlHehl97} and \cite{Hehl00}.
A way to derive the Maxwell equations by means of an operational
approach \cite{LaemmerzahlHehlMacias99} introduces the
electromagnetic field $F$ via consideration of charged particle
interferometry.  Since the phase shift for small areas is
found to be proportional to an interferometer's area,
$\phi \sim$ area.  The proportionality factor can be defined to be
the  electromagnetic field $F$: $\phi = \frac{1}{2} F_{\mu\nu}
\sigma^{\mu\nu}$ (the area is described by an antisymmetric second
rank tensor).  In this way we {\it operationally} defined the
electromagnetic field.

The experimental uniqueness of the phase shift immediately requires
\begin{equation}
0 = dF \, . \label{HomMaxwell}
\end{equation}
In a $3+1$--slicing of the manifold, these equations consist in
three dynamical equations $\pounds_v \bar B = \bar d \bar E$ ($v$ is
a ``timelike'' vector field) $\bar d \bar B = 0$
\cite{LaemmerzahlHehlMacias99}. Another three dynamical
equations are needed.

These other equations can be obtained by requiring that the
Maxwell field obey an {\it evolutional structure} with respect to
the slicing.  This is equivalent to requiring an evolution equation
$\pounds_v F =  \lambda(F)$ that is compatible with (\ref{HomMaxwell}).
In coordinates: $\pounds_v F_{\hat\mu\hat\nu} =
\lambda_{\hat\mu\hat\nu}(F)$, $\pounds_v F_{\mu 0} = \lambda_\mu(F)
= \lambda^0_\mu(F)
+ j_\mu$, where spatial indices are denoted by a hat.  The first
equation
is contained in (\ref{HomMaxwell}).
We call the part of $\lambda_\mu(F)$ which remains as $F
\rightarrow 0$ the {\it source} $j$ of electromagnetic field.
The requirement that the {\it superposition principle} should hold
leads to a linear evolution equation as well as to constraints
which are fulfilled by the superposition, too.

The additional requirement that the electromagnetic field should
propagate with a {\it finite propagation speed} implies that
$\lambda$ is a differential operator and that it should be of first
order only \cite{IvriiPetkov74}.
That gives our generalized Maxwell equations the form
\begin{equation}
\partial_{[\rho} F_{\mu\nu]} = 0 \, \qquad
\partial_\nu (\lambda_\mu^{\phantom{\mu}\nu\rho\sigma}
F_{\rho\sigma}) + \bar\lambda_\mu^{\phantom{\mu}\rho\sigma}
F_{\rho\sigma} = j_\mu \, .\label{eq:GenMaxwell2}
\end{equation}
In general, these equations show birefringence, that is, two light
cones, which of course violate Local Lorentz Invariance (LLI). The  
requirement of a {\it unique
light cone} implies the existence of a non--degenerate second rank
tensor $g^{\mu\nu}$ so that $\lambda^{\mu\nu\rho\sigma}$ can be
rewritten as \cite{Ni73}
\begin{equation}
\lambda^{\mu\nu\rho\sigma} = \frac{1}{2} \sqrt{- g} (g^{\rho[\mu}
g^{\nu]\sigma} - g^{\sigma[\mu} g^{\nu]\rho}) + \theta
\epsilon^{\mu\nu\rho\sigma}
\end{equation}
where $\epsilon^{\mu\nu\rho\sigma}$ is the total antisymmetric
symbol.
Therefore we can rewrite the inhomogeneous Maxwell equations
(\ref{eq:GenMaxwell2}) as ($D$ is the Riemannian covariant derivative)
\begin{equation}
j^\nu = D_\mu F^{\mu\nu} + \Sigma^\nu_{\mu\rho}
F^{\mu\rho} \qquad \text{with}\quad \Sigma^\nu_{\mu\rho} :=
\bar\lambda^\nu_{\mu\rho} - \epsilon^{\nu\sigma}{}_{\mu\rho}
\partial_\sigma\theta - \left\{{\phantom{|}}^\sigma_{\sigma
[\mu}\right\}
\delta^\nu_{\rho]} \, .
\end{equation}
The form of $\Sigma$ can be restricted by requiring {\it
charge conservation} in the form $D_\mu j^\mu = 0$.
The most general $\Sigma$ compatible with this requirement is
\begin{equation}
\Sigma_{\mu\nu\rho} = \epsilon_{\mu\nu\rho}{}^\sigma
\partial_\sigma \theta \, .
\end{equation}
Ni \cite{Ni77}  has shown that this coupling does not
violate the weak equivalence principle for falling charges.
In addition, it was established in \cite{deSabbataGasperini81} that
vacuum polarization effects of the Dirac equation in a space--time
with torsion leads to an effective Maxwell equation with an
additional coupling of the above form.
$\theta$ can be interpreted as {\it axion} field which appears in
the  low--energy limit of string theory and is a candidate for the
dark matter in the universe.
These effective Maxwell equations can be obtained from a Lagrangian
\cite{PuntigamLaemmerzahlHehl97}.

The coupling to $\theta$ results in a violation of Local Position
Invariance. However, in a rotating and accelerating frame with rotation  
$\mbox{\boldmath$\omega$}$ and acceleration $\mbox{\boldmath$a$}$
(stationary situation, Newtonian potential $U(0) = 0$) the above Maxwell  
equations acquire the form (with the PPN parameter $\gamma$)
\begin{eqnarray}
j^0 & = & \mbox{\boldmath$\nabla$} \cdot \mbox{\boldmath$E$} -
{(\mbox{\boldmath$a$} - (1 + \gamma) \mbox{\boldmath$\nabla$} U)}
\cdot \mbox{\boldmath$E$} - 2 {(\mbox{\boldmath$\omega$} -
\mbox{\boldmath$\nabla$}\theta)} \cdot\mbox{\boldmath$B$}
\nonumber\\
\mbox{\boldmath$j$} & = & - \mbox{\boldmath$\nabla$} \times
\mbox{\boldmath$B$} - {(\mbox{\boldmath$a$} - (\gamma + 1)
\mbox{\boldmath$\nabla$} U)} \times\mbox{\boldmath$B$} + 2
{(\mbox{\boldmath$\omega$} - \mbox{\boldmath$\nabla$}\theta)}
\times\mbox{\boldmath$E$}
\end{eqnarray}
which means that gravity and the pseudoscalar field can be
transformed away.
The effects of $\mbox{\boldmath$\nabla$}\theta$ or axial torsion
can be simulated by a rotation of the reference frame
\cite{vonderHeyde75}.

The coupling to $\mbox{\boldmath$\nabla$}\theta$ amounts to a
non--metric Faraday--effect: If at the worldline of the source the
polarization of the electromagnetic field is parallelly propagated,
$D_u F = 0$, then the observer sees a rotation of the
polarization along his worldline, $D_v F_{\mu\nu} = 2 \, \dot x \,
k^\rho (\partial_\tau\theta) \,
\epsilon_{\rho[\mu}^{\phantom{\rho[\mu}\sigma\tau} F_{\nu]\sigma}$.
The absence of such an effect implies usual
Maxwell equations coupled to a space--time metric only:
\begin{equation}
d F = 0 \qquad\qquad d * F = j  \, .
\end{equation}

\section{Experiments testing the Maxwell equations}

In principle, observation of any electromagnetic phenomenon can provide
a basis for a test of the validity of the Maxwell equations, but
phenomena associated with wave propagation are particularly
well suited to imposing sharp constraints on the dynamics of the
electromagnetic field since astronomical observations can sometimes
impose limits on effects that have built up in the course of
propagation over huge distances.  Laboratory tests can be considered
when circumstances do not allow astronomical observations to distinguish
between effects caused by departures from familiar Maxwell dynamics and
by other physical processes.

\subsection{Tests of the generalized Maxwell equations}

First we derive a wave equation from the generalized Maxwell
equations (\ref{eq:GenMaxwell2})
\begin{equation}
\partial_{[\mu} j_{\nu]} = \delta_{[\mu}^\alpha
\lambda_{\nu]}{}^{\kappa\rho\sigma} \partial_\alpha \partial_\kappa
F_{\rho\sigma} + \widetilde\lambda_{[\nu\mu]}{}^{\kappa\rho\sigma}
\partial_\kappa F_{\rho\sigma} +
\widehat\lambda_{[\nu\mu]}{}^{\rho\sigma} F_{\rho\sigma} \, ,
\label{GenWaveEquation}
\end{equation}
where $\widetilde\lambda$ and $\widehat\lambda$ are combinations of
the original $\lambda$ and $\bar\lambda$ and its derivatives.
If $\lambda_\mu{}^{\nu\rho\sigma} = \frac{1}{2} (\delta_\mu^\rho
g^{\nu\sigma} - \delta_\mu^\nu g^{\rho\sigma}) +
\delta\lambda_\mu{}^{\nu\rho\sigma}$ and if the derivatives of the
coefficients are assumed to be negligible, then we get from
(\ref{GenWaveEquation}) for $j = 0$
\begin{equation}
0 = \left(\delta_\mu^\rho \delta_\nu^\sigma \square + 2 \delta
\lambda_{[\mu}{}^{\kappa\rho\sigma} \partial_{\nu]}
\partial_\kappa\right) F_{\rho\sigma} + 2
\bar\lambda_{[\mu}{}^{\rho\sigma} \partial_{\nu]} F_{\rho\sigma} \,
.  \label{GenWaveEquation2}
\end{equation}
We assume that $\delta\lambda_\nu{}^{\nu\rho\sigma}$ as well as
$\bar\lambda_\nu{}^{\rho\sigma}$ are small.

\subsubsection{Propagation of characteristics}

First we discuss the behavior of characteristics, or shock waves,
predicted by (\ref{GenWaveEquation2}).  This is determined by this
equation's principal polynomial
($k$ is the normal of the characteristic surface)
\begin{equation}
0 = \det\left(\delta_{[\mu}^\rho \delta_{\nu]}^\sigma k^2 + 2
\delta \lambda_{[\mu}{}^{\kappa\rho\sigma} k_{\nu]} k_\kappa\right)
\, .
\end{equation}
This prediction and the results of astrophysical observations that
search for  birefringence effects impose the constraint $\delta
\lambda_{\mu}{}^{\kappa\rho\sigma} \leq 10^{-28}$
\cite{HauganKauffmann92}.

\subsubsection{Propagation of plane waves}

If we take the electromagnetic field to be a plane wave $F = F_0
e^{- i k x}$, and do not take the eikonal limit, then the
equation (\ref{GenWaveEquation2}) implies
\begin{equation}
0 = \det\left(\delta_{[\mu}^\rho \delta_{\nu]}^\sigma k^2 + 2
\delta \lambda_{[\mu}{}^{\kappa\rho\sigma} k_{\nu]} k_\kappa + 2 i
\bar\lambda_{[\mu}{}^{\rho\sigma} k_{\nu]}\right) \, .
\end{equation}
The factor $i$ associated with the $\bar\lambda$ stems from a term
involving a single field derivative.
To first order in $\delta\lambda$ and $\bar\lambda$ the
corresponding dispersion  relation is
($\det(1 + \Lambda) = 1 + \hbox{tr}\Lambda + {\cal
O}(\Lambda^2)$)
\begin{equation}
0 = (k^2)^5 \left(k^2 + \delta \lambda_{[\mu}{}^{\kappa\mu\nu}
k_{\nu]} k_\kappa + \bar\lambda_{[\mu}{}^{\mu\nu} k_{\nu]}\right) +
{\cal O}(\delta\lambda^2, \bar\lambda^2) \, .
\end{equation}
Non--trivial solutions are given by ($k^2 = \omega^2 - {\rm k}^2$)
\begin{eqnarray}
\omega_{1,2} & = & - \frac{1}{2} ((\delta \lambda_{\mu}{}^{0\mu
\hat \nu} + \delta \lambda_{\mu}{}^{\hat\kappa \nu 0}) k_{\hat \nu}
+ i \bar\lambda_{\mu}{}^{\mu 0}) \nonumber\\
& & \pm \left({\rm k}\left(1 - \frac{1}{2} \delta
\lambda_{\mu}{}^{0\mu 0}\right) - \frac{1}{2} \left(\delta
\lambda_{\rho}{}^{\hat\mu \rho \hat \nu} \frac{k_{\hat\mu}}{{\rm k}}
k_{\hat\nu} + i \bar\lambda_{\mu}{}^{\mu \hat \nu} \frac{k_{\hat
\nu}}{{\rm k}}\right) \right) \, .
\end{eqnarray}
Note the presence of two effects: (i) an anisotropic relation between
$k_{\hat\mu}$ and $\omega$ due to $\delta\lambda$, indicating a
violation of LLI, and (ii) imaginary
terms due to a trace of $\bar\lambda$, indicating a damping of
a plane wave's intensity.  This damping is independent of wave length
but depends on the direction in which a wave propagates.
Knowledge of the distance and intrinsic brightness of stars should
permit the estimation of the strength of any such damping.
In principle, tests of LLI can lead to estimates of $\delta\lambda$
and observations of damping lead to estimates of $\bar\lambda$ and,
thus, constraints on charge non-conservation.  Recall that charge
conservation forced $\bar\lambda=0$.

In the case of a coupling to the axion alone we find the exact
dispersion relation $\omega = \pm {\rm k} \sqrt{1 \pm \dfrac{1}{{\rm
k}} \dfrac{\partial\theta}{\partial t}}$
so that the observed frequencies are $\omega = k \pm
\dfrac{\partial\theta}{\partial t} - \dfrac{1}{2 {\rm k}}
\left(\dfrac{\partial\theta}{\partial t^\prime}\right)^2 + {\cal
O}({\rm k}^{-2})$, that is, waves with opposite helicity
propagate with different speeds.  The results of astronomical
observations lead to the constraint
$\dfrac{\partial \theta}{\partial t} \leq 10^{-32}\;\hbox{eV}$
\cite{CarrolFieldJackiw91}.

Predicted effects of a hypothetical $\theta$ coupling on energy levels
of atoms and existing atomic physics data imply the constraint
$\nabla\theta \leq  10^{-8}\;{\hbox{m}}^{-1}$ \cite{LaemmerzahlNi99}.

\subsection{Test of well--posedness of Cauchy--problem}

If for an evolution the Cauchy--problem is not well posed, then
this evolution depends on its history or possesses memory.
In mathematical terms this means, that, under certain
circumstances, one has to pose all time derivatives up to infinite
order: $F$, $\partial_t F$, $\partial_t^2 F$, ..., $\partial_t^n F$,
... .
In a first approximation one may ask, whether there is any need to
pose also $\partial_t F$ in addition to $F$.
For the usual Maxwell equations (here we assumne, for simplicity, a metric)  
we therefore have to add a term
$\partial_t^2 F$:
\begin{equation}
\tilde a_\mu{}^{\rho\sigma} \partial_0^2 F_{\rho\sigma} +
\partial_\nu F_\mu{}^\nu = j_\mu \, .
\end{equation}
The addition of such a term clearly violates LLI.
In addition, if such a term is present, the homogeneous Maxwell
equations reduce to constraints, because they are of first order
only.

This term can be analyzed by the propagation of plane waves.
The dispersion relation is
\begin{eqnarray}
0 & = & \det\left(k^2 \delta_{[\mu}^{[\rho} \delta_{\nu]}^{\sigma]}
+ \tilde a_{[\mu}{}^{[\rho\sigma]} k_{\nu]} \omega^2\right)
\nonumber\\
& \approx & (k^2)^5 \left(\omega^2 - {\rm k}^2 + \frac{1}{2}
\omega^2 \left(\tilde a_{0}{}^{[0\hat\nu]} k_{\hat\nu]} + \tilde
a_{\hat\mu}{}^{[\hat\mu 0]} \omega + \tilde
a_{\hat\mu}{}^{[\hat\mu\hat\nu]} k_{\hat\nu}\right) \right)
\end{eqnarray}
leading to the non--trivial solutions
\begin{eqnarray}
\omega_{1,2} & = & \pm {\rm k} \left(1 \mp \frac{i}{8} (\tilde
a_{0}{}^{[0\hat\nu]} + \tilde a_{\hat\mu}{}^{[\hat\mu\hat\nu]})
k_{\hat\nu} - \frac{i}{8} \tilde a_{\hat\mu}{}^{[\hat\mu 0]} {\rm
k}\right) \\
\omega_3 & = & \frac{i}{\tilde a_{\hat\mu}{}^{[\hat\mu 0]}} \left(2
+ i \frac{1}{2} (\tilde a_{0}{}^{[0\hat\nu]} + \tilde
a_{\hat\mu}{}^{[\hat\mu\hat\nu]}) k_{\hat\nu} + \frac{1}{8} (\tilde
a_{\hat\mu}{}^{[\hat\mu 0]})^2 {\rm k}^2\right) \, .
\end{eqnarray}
The third solution has no physical relevance because it diverges
for vanishing $\widetilde a$.
From the first two solutions we conclude that a second
time--derivative in the Maxwell equations will lead to an
anisotropic damping of
plane waves.

\subsection{Test of finite propagation speed}

As for the heat or the Schr\"odinger equation, infinite propagation
speed means that the order of spatial derivatives is larger than the
the order of the time derivative.
In our case this means that we have as modified Maxwell equations
\begin{equation}
\partial_\nu F_\mu{}^\nu + \tilde b_\mu{}^{\hat\kappa \hat\lambda
\rho\sigma} \partial_{\hat\kappa} \partial_{\hat\lambda}
F_{\rho\sigma} = j_\mu \, ,
\end{equation}
which again violates LLI.
Again, these equations can be analyzed through propagation phenomena.
We get as dispersion relation
\begin{eqnarray}
0 & = & \det\left(k^2 \delta_{[\mu}^{[\rho} \delta_{\nu]}^{\sigma]}
+ k_{[\mu} \tilde b_{\nu]}{}^{\hat\kappa\hat\lambda [\rho\sigma]}
k_{\hat\kappa} k_{\hat\lambda}\right) \nonumber\\
& \approx & (k^2)^5 \left(\omega^2 - {\rm k}^2 + \frac{1}{2}
\left(\tilde b_{\hat\mu}{}^{\hat\kappa\hat\lambda [\hat\mu 0]}
\omega + \left(\tilde b_{0}{}^{\hat\kappa\hat\lambda [0\hat\mu]} +
\tilde b_{\hat\nu}{}^{\hat\kappa\hat\lambda [\hat\mu\hat\nu]}\right)
k_{\hat\mu}\right) k_{\hat\kappa} k_{\hat\lambda}\right)
\end{eqnarray}
with the solutions
\begin{equation}
\omega_{1,2} = \pm {\rm k} - \frac{1}{4} \tilde
b_{\hat\mu}{}^{\hat\kappa\hat\lambda [\hat\mu 0]} k_{\hat\kappa}
k_{\hat\lambda} \mp \frac{1}{4 {\rm k}} \left(\tilde
b_{0}{}^{\hat\kappa\hat\lambda [0\hat\mu]} + \tilde
b_{\hat\nu}{}^{\hat\kappa\hat\lambda [\hat\mu\hat\nu]}\right)
k_{\hat\mu} k_{\hat\kappa} k_{\hat\lambda}
\end{equation}
showing up an anisotropic dispersive propagation thus violating LLI.

\section{Conclusion}

We have presented a new test theory for the dynamics of the
electromagnetic field.
This theory was applied to propagation phenomena well
suited to testing the Maxwell equations with astronomical data.
We reviewed existing constraints on conventional anomalous
electromagnetic field dynamics and also discussed the effects
of anomalies associated with the appearance of higher-order field
derivatives in the generalized Maxwell equations.  These latter
anomalies cause either anisotropic wave propagation or wave
damping.  The structure of the test theory reviewed here is
sufficiently general to provide a context for the design and
interpretation of experimental tests of special relativity and
of metric gravitation theories like general relativity.

\vspace*{0.25cm} \baselineskip=10pt{\small \noindent C.L. thanks
the Optikzentrum of the University of Konstanz for financial
support. }

\end{document}